\documentclass[aps,prl,twocolumn,superscriptaddress,showpacs]{revtex4}
\usepackage{graphicx}
\usepackage{color}
\usepackage{bm}
\usepackage{braket}

\usepackage[normalem]{ulem}

\begin{document}

\title{Coulomb correlations intertwined with spin and orbital excitations in LaCoO$_3$}

\author{K. Tomiyasu}
\altaffiliation [email: ] {\emph{tomiyasu@m.tohoku.ac.jp}} 
\affiliation{Department of Physics, Tohoku University, Aoba, Sendai 980-8578, Japan}

\author{J. Okamoto}
\affiliation{National Synchrotron Radiation Research Center, Hsinchu 30076, Taiwan}

\author{H. Y. Huang}
\affiliation{National Synchrotron Radiation Research Center, Hsinchu 30076, Taiwan}

\author{Z. Y. Chen}
\affiliation{Department of Physics, National Tsing Hua University, Hsinchu 30013, Taiwan}

\author{E. P. Sinaga}
\affiliation{Department of Physics, Tohoku University, Aoba, Sendai 980-8578, Japan}

\author{W. B. Wu}
\affiliation{National Synchrotron Radiation Research Center, Hsinchu 30076, Taiwan}

\author{Y. Y. Chu}
\affiliation{National Synchrotron Radiation Research Center, Hsinchu 30076, Taiwan}

\author{A. Singh}
\affiliation{National Synchrotron Radiation Research Center, Hsinchu 30076, Taiwan}

\author{R.-P. Wang}
\author{F. M. F. de Groot}
\affiliation{Inorganic Chemistry and Catalysis, Utrecht University, Universiteitsweg 99, 3584 CG Utrecht, The Netherlands}

\author{A. Chainani}
\affiliation{National Synchrotron Radiation Research Center, Hsinchu 30076, Taiwan}

\author{S. Ishihara}
\affiliation{Department of Physics, Tohoku University, Aoba, Sendai 980-8578, Japan}

\author{C. T. Chen}
\affiliation{National Synchrotron Radiation Research Center, Hsinchu 30076, Taiwan}

\author{D. J. Huang}
\altaffiliation [email: ] {\emph{djhuang@nsrrc.org.tw}} 
\affiliation{National Synchrotron Radiation Research Center, Hsinchu 30076, Taiwan}
\affiliation{Department of Physics, National Tsing Hua University, Hsinchu 30013, Taiwan}

%\affilliation[$^\dagger$]{These authors contributed equally to this work.}
%\affilliation[$^\ast$]{Corresponding author. E-mail: tomiyasu@m.tohoku.ac.jp (K.T.); djhuang@nsrrc.org.tw (D.J.H.)}

%\keywords{Keyword1, Keyword2, Keyword3}
\date{\today}
\begin{abstract}
We carried out temperature-dependent (20 - 550 K)
measurements of resonant inelastic X-ray scattering on LaCoO$_3$
to investigate the evolution of its electronic structure across the spin-state crossover.
In combination with charge-transfer multiplet calculations, we accurately quantified the renomalized crystal-field excitation energies and spin-state populations. We show that the screening of the on-site Coulomb interaction of $3d$ electrons is orbital selective and coupled to the spin-state crossover in LaCoO$_3$. The results establish that the gradual spin-state crossover is associated with a relative change of Coulomb energy versus bandwidth, leading to a Mott-type insulator-to-metal transition.
\end{abstract}


%pacs
%71.30.+h	Metal-insulator transitions and other electronic transitions
%71.70.Ch	Crystal and ligand fields
%75.10.Dg	Crystal-field theory and spin Hamiltonians (see also 71.70.Ch Crystal and ligand fields)
%75.30.Wx	Spin crossover
%75.25.-j	Spin arrangements in magnetically ordered materials (including neutron and spin-polarized electron studies, synchrotron-source x-ray scattering, etc.) (for devices exploiting spin polarized transport, see 85.75.-d) 
%75.70.Tj	Spin-orbit effects (see also 71.70.Ej Spin-orbit coupling, Zeeman and Stark splitting, Jahn-Teller effect)
%78.70.Ck	X-ray scattering
%78.70.En	X-ray emission spectra and fluorescence

\flushbottom
\maketitle
\thispagestyle{empty}
The orbital degree of freedom of an electron characterizes the shape of the electron cloud and its wave function. It plays an essential role in the physics of phase transitions in solids via the coupling of charge, spin and lattice degrees of freedom, even in the presence of strong Coulomb interactions, for example, as in Mott insulators. The spatial redistribution of the electron cloud as a function of an external parameter such as  temperature often manifests as co-operative phenomena leading to a metal-insulator transition \cite{Imada98}, orbital ordering \cite{Tokura00,Kugel73}, nematic transition \cite{Baek15,Watson15}, spin-state transition \cite{Gutlicg04,Hohenberger12,Krewald16,Nomura11,Ohkoshi11}, etc. These results in exotic properties like superconductivity, quantum criticality, colossal magnetoresistance, etc. As the Coulomb interaction is a key to Mott physics \cite{Medici14,Jakobi13,Medici09,Werner07}, one fundamental question in correlated electron systems with orbital degrees of freedom is: how do the Coulomb correlations change dynamically through the rearrangement of the electronic distribution? This is usually beyond the scope of even multi-orbital model Hamiltonians in which the Coulomb interaction parameters are considered inflexible. An important theoretical advance in this direction  is the role of orbital selective screening \cite{Solovyev96}. The effective Coulomb interaction for $t_{2g}$ electrons was shown to be significantly reduced due to efficient $e_{g}$ electron screening, providing an improved understanding of LaMO$_{3}$ (M = $3d$ transition metals from Ti to Cu) series of perovskite oxides. This concept of the screened on-site Coulomb interaction has been developed recently using the constrained random-phase-approximation technique \cite{Aryasetiawan06,Solovyev08,Imada10}.

In this $Letter$, we exploited resonant inelastic X-ray scattering (RIXS)  to investigate spin-orbital excitations in LaCoO$_3$ and to measure its spin-state populations and the renormalized crystal-field excitations across the spin-state transition. The results indicate that LaCoO$_3$ is an ideal candidate to examine the role of orbital selective screening of Coulomb interactions as a function of temperature.  We found that the spin-state crossover is driven by the thermal excitation of high-spin (HS) states and accompanied by the reduction in effective Coulomb energy and an increase of covalency, culminating in an effective Coulomb-energy-vs-bandwidth type insulator-to-metal transition.

\begin{figure}[t]
\centering
\includegraphics[width=0.95\columnwidth]{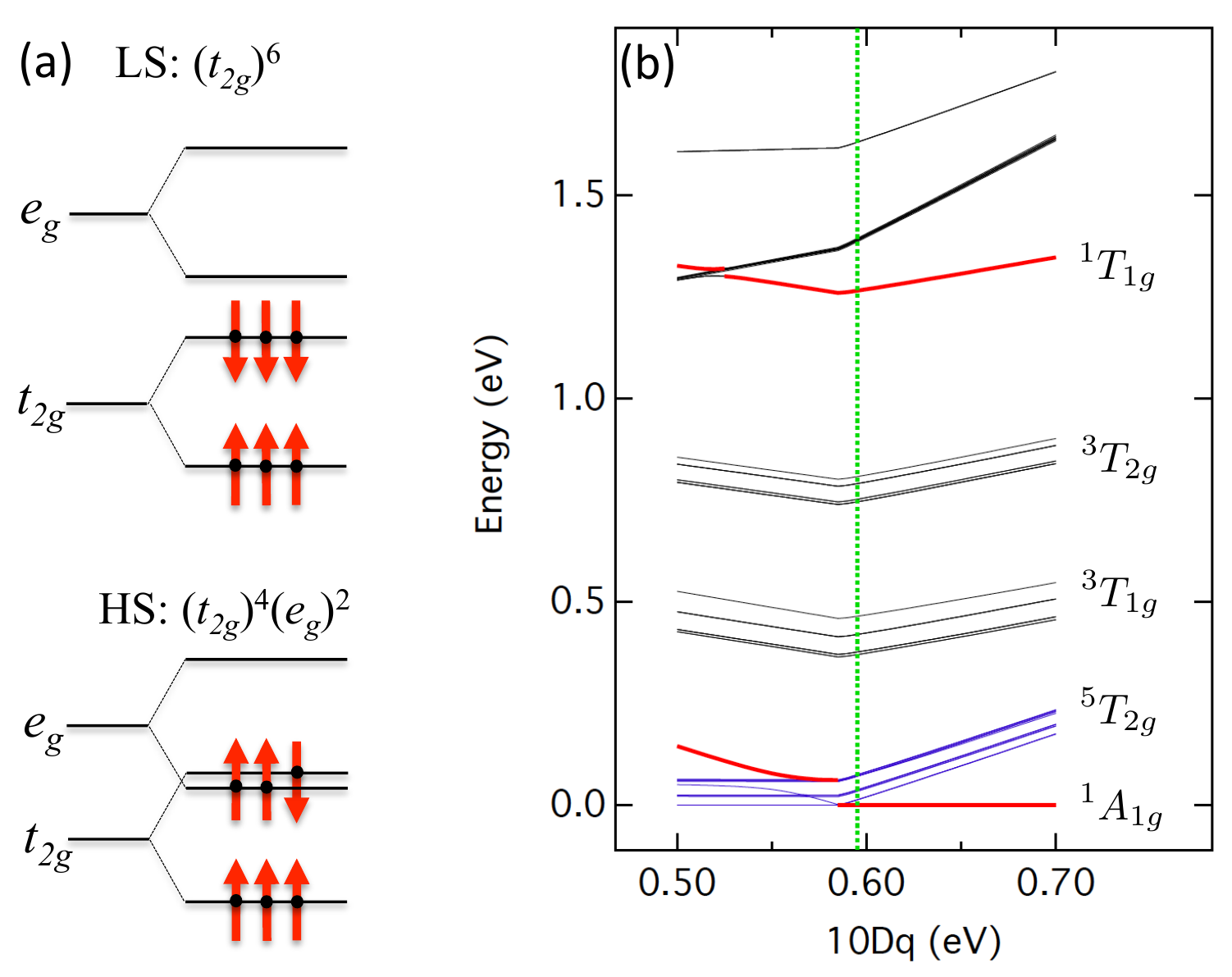}
\caption{Calculated energy diagrams of Co$^{3+}$. (a) Illustration of the one-electron energy diagrams of Co$^{3+}$ in LS and HS states without including the effect of charge transfer between Co $3d$ and O $2p$. (b) Calculated energies of the ground state and excited states of  Co$^{3+}$ plotted with respect to the ground state as a function of $10Dq$. See Ref. \cite{para} for the calculation parameters. The vertical dotted line denotes $10Dq = 0.595$~eV at which the contribution of the $3d^{6}$ configuration in the ground state $^{1}A_{1g}$ is 39.3\%, whereas those of $3d^{7}\underline{L}$ and  $3d^{8}\underline{L}^2$ are, respectively,  50.7\% and  10\%, where $\underline{L}$ denotes a ligand hole.}
\end{figure}

Spin-state  transitions or crossovers between low-spin (LS) and HS states occur in diverse materials \cite{Gutlicg04,Hohenberger12,Krewald16,Nomura11,Ohkoshi11}. LaCoO$_3$ is a prototypical example of spin-state transition in solids \cite{Cambi31,1967Raccah,1953Jonker}. Whether the Co$^{3+}$ in LaCoO$_3$ is in a LS or HS configuration is determined by a competition between the intra-atomic Hund's exchange energy $J_H$ and the crystal field splitting $10Dq$ as illustrated in Fig. 1(a).  The calculated energy level diagram of Co$^{3+}$ as a function of $10Dq$ shown in Fig. 1(b) demonstrates that, for a large $10Dq$, the electronic configuration energetically favours the LS state, whereas the HS state is favoured by an increased $J_H$, or a decreased $10Dq$ or hybridisation \cite{para}. LaCoO$_3$ undergoes two electronic crossovers in the temperature range between 30 and 600~K. For temperatures below 30~K, LaCoO$_3$ is undisputedly identified as being in a LS state. Its magnetic susceptibility $\chi(T)$ rises sharply with increasing temperature and exhibits a maximum about 100~K, referred to as the spin-state transition \cite{1996Yamaguchi,1967Raccah}. 
The second crossover, often referred to as a metal-insulator transition, is at 530~K where the heat capacity shows a maximum \cite{Stolen97}.

The temperature dependence of $\chi(T)$ describing the spin-state transition of LaCoO$_3$ was originally interpreted as implying a gradual population increase of HS states with a fixed activation energy, an energy required to excite the ground state to the first excited state \cite{1967Raccah}. This scenario however led to an overestimated $\chi(T)$ and motivated an intermediate-spin (IS) description. Band-structure calculations with Coulomb correlations included gave a strong boost for the IS picture \cite{Korotin96,Saitoh97,Radaelli02,Maris03,Pandey08,Doi14}. In contrast, electron spin resonance \cite{Noguchi02}, inelastic neutron scattering \cite{Podlesnyak06},  and X-ray absorption spectroscopy (XAS) \cite{Haverkort06} showed that the lowest-energy excited state is a HS state which exhibits additional splitting owing to the spin-orbit interaction. To explain results of specific heat and XAS in terms of the HS-LS scenario, one needs to adopt a strong temperature-dependent increase of the crystal field \cite{Kyomen03,Kyomen05,Eder10,Haverkort06}. For example, the XAS work explained
the transition using a temperature-dependent increase of activation energy \cite{Haverkort06}. However this leads to a puzzle \cite{Eder10}: for the LS ground state, an increased activation energy implies an increased bare ionic crystal-field splitting $10Dq$, inconsistent with a reduction in $10Dq$ expected from the experimentally known expansion in Co-O bond lengths \cite{Radaelli02}. 
This puzzle points to an important issue regarding how the bare $10Dq$ and the Coulomb correlation are modified via the change in $t_{2g}$ and $e_g$ orbital occupancy. 

%Experimental
Using the AGM-AGS spectrometer \cite{Lai2014} at beamline TLS 05A1 of the National Synchrotron Radiation Research Center (NSRRC) in Taiwan, we measured RIXS on a polycrystalline LaCoO$_{3}$ and a single-crystal LaCoO$_{3}$(001) samples \cite{sample}  at incident photon energies set to specific energies about the $L_3$ ($2p_{3/2}\rightarrow 3d$) X-ray absorption edge of Co.  The scattering angle defined as the angle between the incident and the scattered X-rays was $90^\circ$, and the incident angle from the crystal $ab$ plane was $20^\circ$. The polarization of the incident X-ray was switchable between $\pi$ and $\sigma$  polarizations, i.e. the polarization within and perpendicular to the scattering plane, respectively, and the polarization of scattered X-rays was not analyzed. The energy bandwidth of the incident X-rays was 500~meV and the total RIXS energy resolution was 80~meV because the energy compensation method was used to ensure a high-resolution measurement in the  energy loss scheme \cite{Lai2014}. The beam diameter of incident X-ray at the sample was about 0.5~mm. 

Unlike X-ray absorption, RIXS probes electronic excitations such as orbital and spin excitations without the presence of a $2p$ core hole \cite{Ament2011,Schlappa12,Lai2014}. RIXS enables a complete characterisation of electronic excitations derived from different spin channels and quantifies spin-state populations. Figure 2 illustrates how RIXS measures the crystal-field excitation from the $t_{2g}$ to the $e_{g}$ bands, and shows the spectra obtained from LaCoO$_{3}$ excited by various incident photon energies at 20~K.  A fluorescence-like RIXS feature initially develops and overlaps with the $d$-$d$ excitation feature near 1.27 eV when the incident energy is set to 0.5~eV below the energy of maximum XAS intensity \cite{XAS}. Such a broad RIXS feature arises from the continuum of particle-hole excitations; its energy shifts with the incident X-ray energy.  Two Raman-like features of energy losses centerd at 0.6~eV and 1.27~eV appear in the RIXS spectra.  The excitations from the LS ground state  to IS states of symmetries $^{3}T_{1g}$ and  $^{3}T_{2g}$ yield the broad 0.6-eV RIXS feature (see Fig. 1(b)). Similarly, the 1.27-eV feature arises from excitation from the LS ground state to another LS state of symmetry $^{1}T_{1g}$.  Of the six $t_{2g}$ electrons, one is promoted to the empty $e_g$ state without spin change. 
The RIXS excitation energy at 1.27~eV, i.e. the renormalized crystal-field excitation energy, can be explained in terms of the eigen-energies from charge-transfer multiplet calculations, as indicated by the green dashed line in Fig. 1(b). 

\begin{figure}[t]
\centering
\includegraphics[width=0.95\columnwidth]{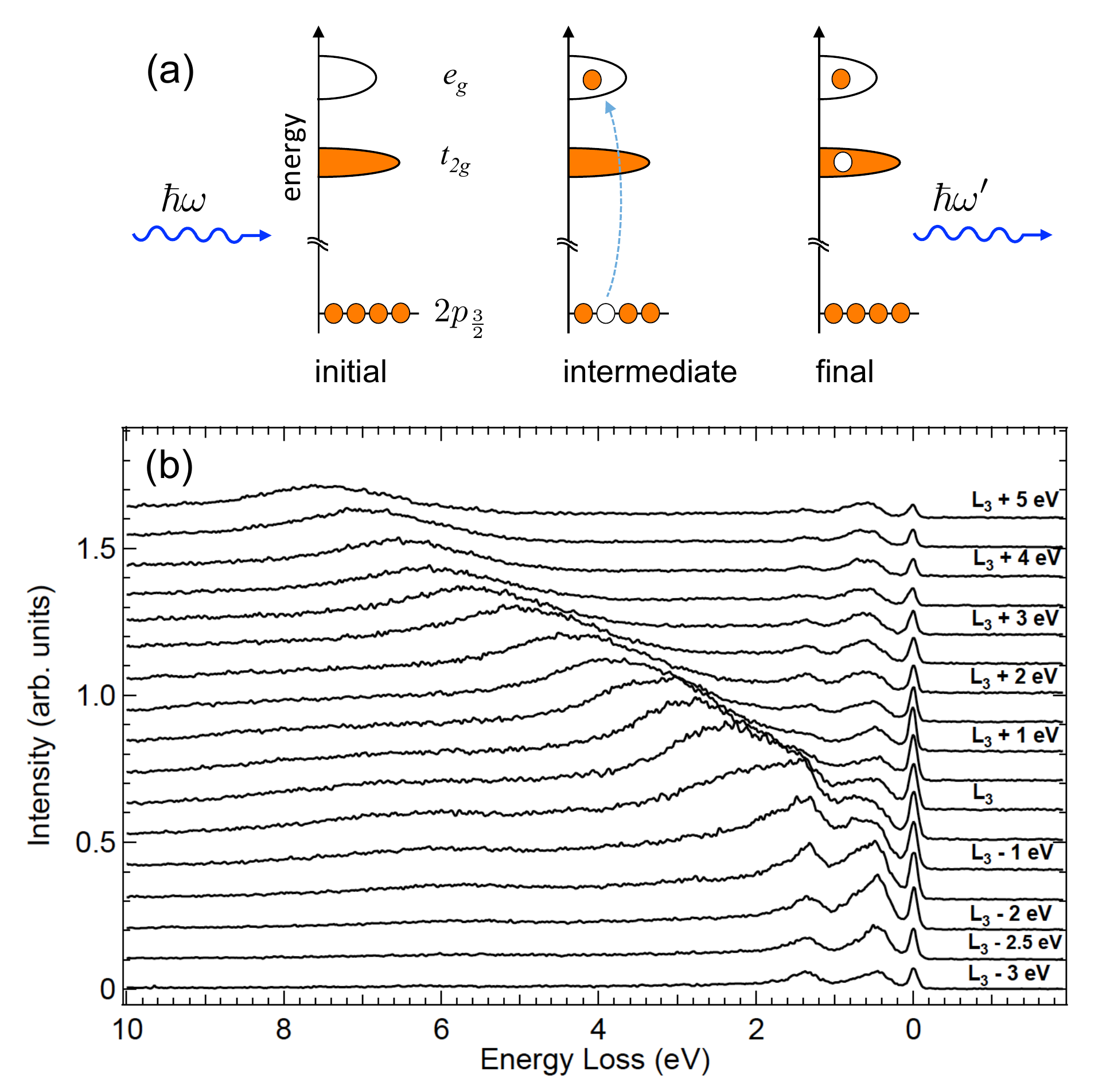}
\caption{RIXS spectra of polycrystalline LaCoO$_3$ recorded with incident X-rays of energy varied across the Co $L_3$ edge. (a) Illustration of a RIXS process of a LS state with the incident and outgoing photons of energies $\hbar\omega$ and $\hbar\omega'$, respectively. (b) RIXS spectra recorded with the sample at temperature 20 K by using incident X-rays of $\sigma$ polarization and with various incident energy varied in steps of 0.5~eV. Spectra are plotted with a vertical offset for clarity.}
\end{figure}

Figure 3(a) shows RIXS spectra for temperatures from 20~K to 550~K. The RIXS features with energy below 2~eV depend strongly on temperature. When the temperature was increased, the intensity of the 1.27-eV excitation decreased. In contrast, a remarkable increase in intensity was observed within 0.2 eV of the elastic peak due to excitations from thermally excited HS states to other HS states within of the same $^{5}T_{2g}$ manifold. Figure 3(b) compares RIXS of three other cobaltates \cite{2004Hu,2003Knee,Kumar14}: EuCoO$_3$, LaCo$_{0.5}$Ni$_{0.5}$O$_3$, and Sr$_2$CoO$_3$Cl, which are considered as reference systems for LS, IS,  and HS ground states, respectively.  
Except for the absolute excitation energies determined by the $3d$ electronic energies including crystal field splitting, exchange energy, covalency and Coulomb energy, the RIXS lineshape of LaCoO$_{3}$ at 20~K resembles that of EuCoO$_{3}$ satisfactorily, confirming that the former is also in a LS state. The observed excitation energies can be explained by calculated RIXS energies shown in Fig. 3(a) for the ground state $^{1}A_{1g}$ as the initial state \cite{para}. Interestingly, the RIXS of LaCoO$_{3}$ at 450~K is similar to that of high-spin Sr$_2$CoO$_3$Cl except for the absolute excitation energies. The above comparison implies that LaCoO$_{3}$ at 450~K is dominated by a HS state, in agreement with the HS scenario \cite{Noguchi02,Podlesnyak06,Haverkort06}. In addition, the RIXS lineshape at 550~K resembles that of LaCo$_{0.5}$Ni$_{0.5}$O$_3$, in which Co$^{3+}$ is in an IS state \cite{Kumar14}. 

\begin{figure}[t]
\centering
\includegraphics[width=0.95\columnwidth]{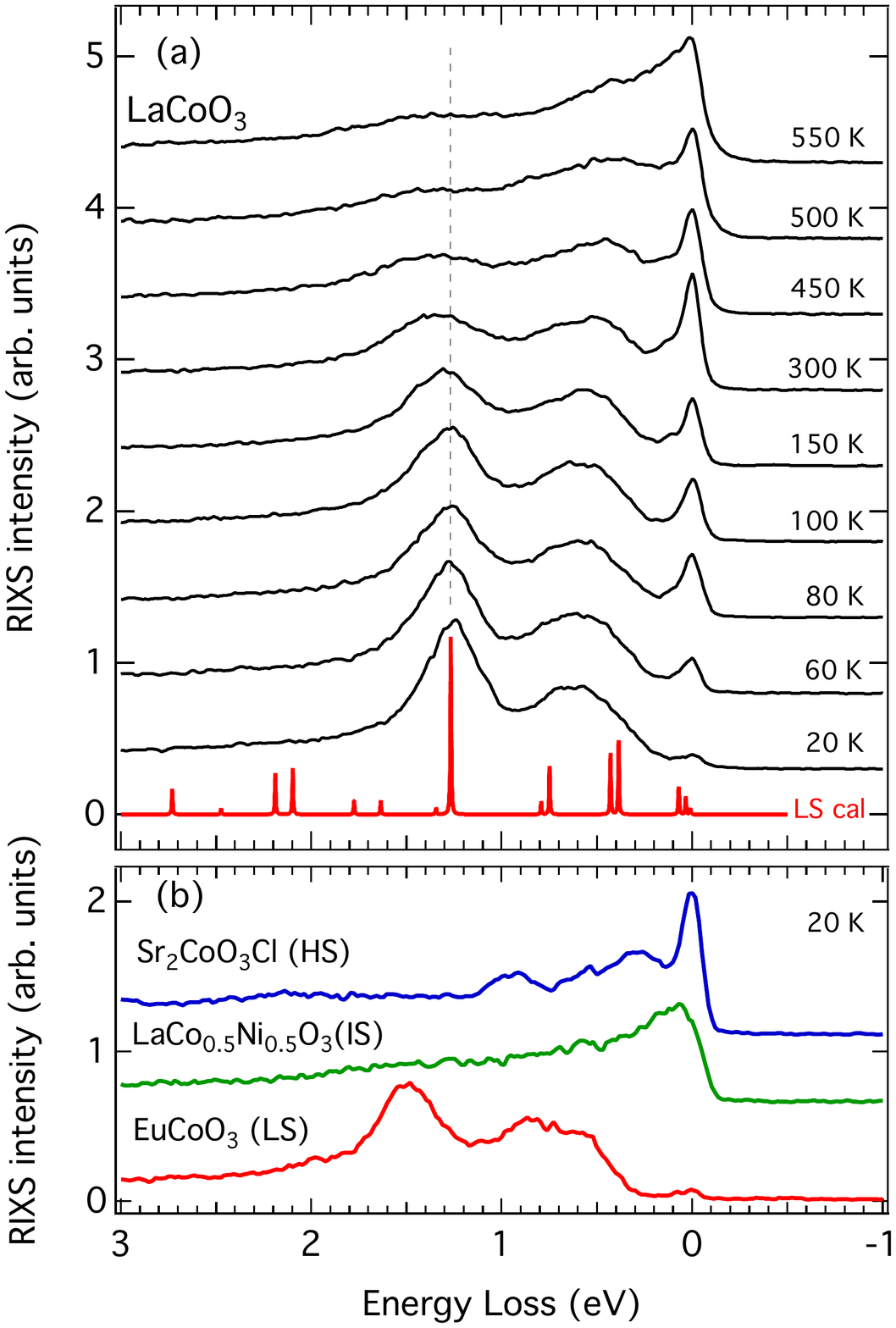}
\caption{Temperature-dependent RIXS measurements. (a) RIXS spectra of single-crystal LaCoO$_{3}$ at varied temperatures.  The spectra have been normalized to the integrated area of the inelastic background between 7 and 10 eV. The red curve (LS cal) shows the calculated RIXS spectral weight of LS Co$^{3+}$ with $10Dq=0.595$~eV; other parameters are the same as those of Fig. 1(b).  The vertical dashed line gives a guide to the eye. (b) RIXS of EuCoO$_{3}$, LaCo$_{0.5}$Ni$_{0.5}$O$_3$, and Sr$_2$CoO$_3$Cl.  All RIXS spectra were recorded with $\pi$-polarized incident X-rays of energy set to $L_{3}-2.5$~eV. Spectra are plotted with a vertical offset for clarity. }
\end{figure}

To characterize the evolutions of the spin-state populations, we analyzed temperature-dependent RIXS by using the linear combination of two reference spectra for LS and HS states. The 20-K spectrum was used for the LS reference spectrum.  As Figs. 3(a) and 3(b) disclose that the RIXS of LaCoO$_{3}$ at 500~K contains a mixture of  HS and IS states, we adopted the 500-K RIXS after a subtraction of the spectral weight contributed by the IS state as the HS reference spectrum \cite{IS}.  Figures 4(a), 4(b) and 4(c) show examples of analysis for 60, 100 and 300 K, respectively.  The obtained combination coefficients provide a measure of the LS and HS populations in the mixed spin state.  With the combination coefficients as free parameters in the fits, the simulations capture fairly well the evolution of the RIXS spectra except for the elastic scattering. The discrepancy near the elastic scattering is attributed to the increase in  the density of states close to the Fermi level.  The obtained combination coefficients for various temperatures are plotted in Fig. 4(d), i.e. open squares and closed circles, in which the error bars include the uncertainty as a result of the assumption on the IS weight  in the initial choice of the HS basis spectrum.

The HS states of $^{5}T_{2}$ symmetry comprise three manifolds of effective angular momenta $J_{\rm eff}$~=~1, 2 and 3. If each manifold of degeneracy $\nu_{i}$ is approximated to an average energy $E_{i}$,  the HS population containing these three manifolds is scaled with $\Sigma^{3}_{i=1}\nu_{i}e^{-E_{i}/k_{B}T}$, where $k_{B}$ is the Boltzmann constant.
To calculate spin-state populations, we adopted $E_{i}$ from charge-transfer multiplet calculations which explain the RIXS spectrum at 20~K as denoted by green circles in the inset of Fig. 4(d). The calculated curves of the HS and LS populations from these energies are depicted by solid and dashed lines in Fig. 4(d), respectively. These curves agree astonishingly well with those deduced from the coefficients of linear combination of RIXS data. This observation lends further support to the LS-to-HS character of the transition. 

The above RIXS results indicate that the lowest activation energy of LaCoO$_3$ is $13{\pm}1$~meV, consistent with values from inelastic neutron scattering \cite{Podlesnyak06,2005Kobayashi}, nuclear magnetic resonance \cite{Kobayashi00}, and electron spin resonance \cite{Noguchi02}. We used the deduced $E_{i}$ to analyze the magnetic susceptibility $\chi(T)$ through a combination of thermal population picture and mean-field approximation \cite{chi}. Figure 4(e) shows the comparison between the measured and calculated $\chi(T)$ curves, which agree satisfactorily with each other.  The averaged $g$-factor is $g=3.1$, consistent with the results obtained from electron spin resonance \cite{Noguchi02}.

\begin{figure}[t]
\centering
\includegraphics[width=0.95\columnwidth]{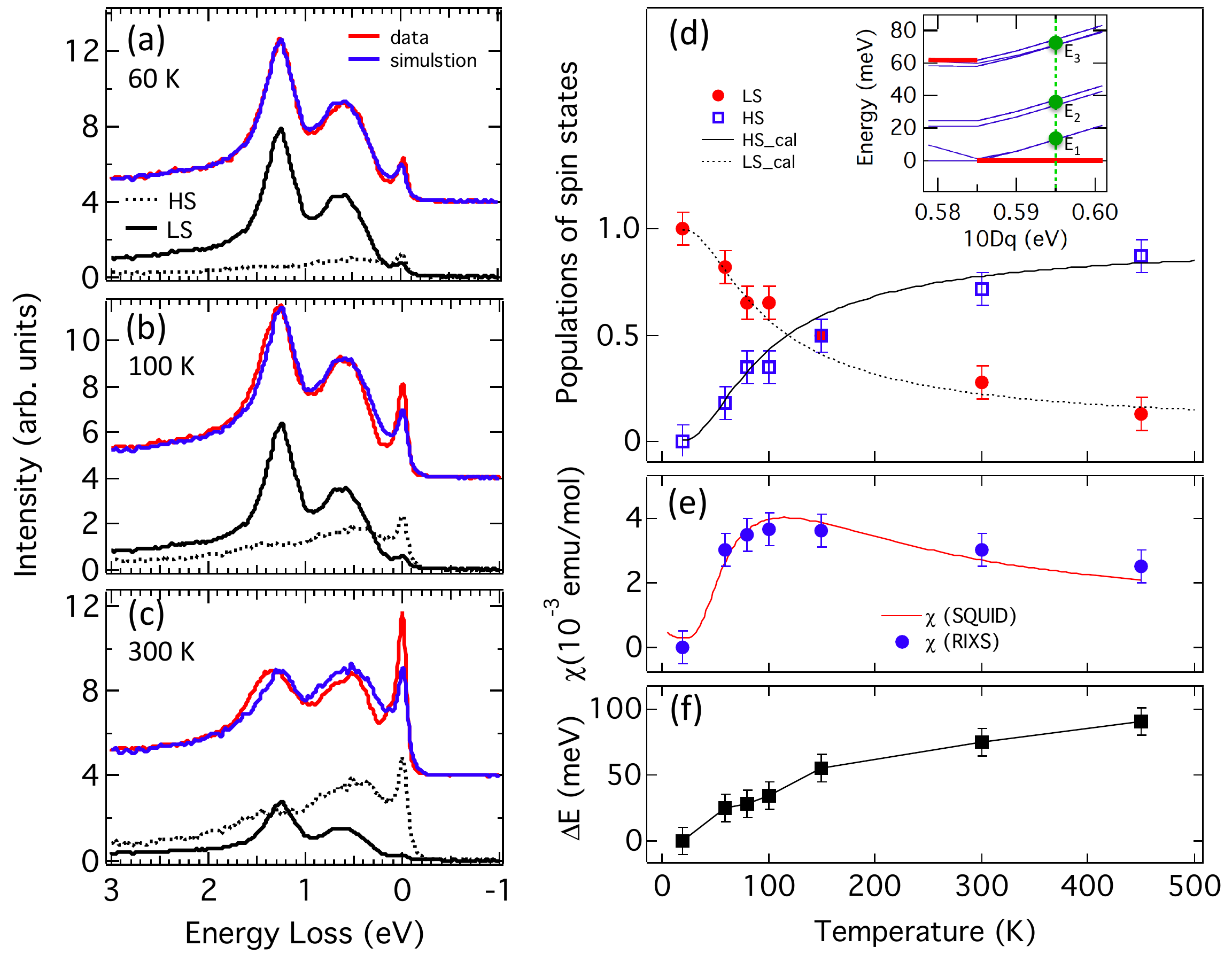}
\caption{Evolution of spin states of LaCoO$_3$ from 20 K to 450 K. (a), (b) \& (c) Simulated RIXS of LaCoO$_3$ at temperatures 60 K, 100 K, and 300 K by using the LS and HS reference spectra discussed in the main text. (d) LS \& HS populations obtained from simulations for various temperatures. Open squares and closed circles are deduced spin-state populations from RIXS data. The solid line plots the calculated HS population $f_{\rm HS}$ with $E_i$= 13, 35.5  \& 72.5~meV and $\nu_i$= 3, 5 \& 7. The dotted line is the LS population $f_{\rm LS}= 1-f_{\rm HS}$. Inset: Calculated energies of the $^{5}T_{2g}$ states about the LS-HS transition.  Green circles indicate the averaged energies of $J_{\rm eff}$= 1, 2, \& 3 at $10Dq$~=~0.595~eV. (e) Comparison between measured $\chi(T)$ from SQUID and deduced $\chi(T)$ from RIXS data. (f) The energy shift $\Delta E$ between the measured and simulated energies of RIXS excitations from the ground state $^{1}A_{1g}$ to $^{1}T_{1g}$ without spin change as a function of temperature.}
\end{figure}

%This blue shift could result from the contribution of the HS state or from the change in the electronic structure of LaCoO$_{3}$ as explained below. 
%the observed shift of the 1.27-eV RIXS feature arises from a subtle change in the electronic structure of LaCoO$_3$ across the spin-state transition.   

In addition to its intensity decrease, the energy of the RIXS excitation from the ground state to the $^{1}T_{1g}$ state shown in Fig. 3(a) appears to exhibit a blue shift with the increase of temperature. 
The comparison between the measured RIXS and the simulated one shown in Figs. 4(a), 4(b) and 4(c) reveals the blue shift of the 1.27-eV RIXS feature.  
Figure 4(f) plots this energy shift $\Delta E$ as a function of temperature. We first examined the effect of lattice expansion on the energy shift of the 1.27-eV RIXS feature by carrying out charge-transfer multiplet calculations. An expansion of lattice yields a reduction in $10Dq$  and $pd\sigma$. The calculations show that, with $J_H$, $U_{dd}$ and $pd\sigma$ fixed, a reduction in $10Dq$ results in a red shift of the 1.27-eV feature. Similarly, a decrease in $pd\sigma$ also causes a red shift  \cite{cal}. Hence this blue shift does not stem from the reduction of the $10Dq$, or from the decrease in covalency due to the thermal expansion of the Co-O bond \cite{Radaelli02}. 

One scenario which could explain the blue shift is the increase in $10Dq$ because of the local contraction of LS CoO$_6$ octahedra in the breathing type Jahn-Teller distortion, whereas the Co-O bond length is expanded on the neighbouring HS sites \cite{Haverkort06}. 
This scenario seemingly explains the blue shift for temperatures below 150~K, at which the HS population is less than 50\%. However, this picture is energetically unfavourable at high temperatures when the two HS ions are nearest neighbours. In addition, the breathing type of lattice distortion was not detected by measurements of extended X-ray absorption fine structure (EXAFS) \cite{Sundaram09}, at least  up to 330~K.  
On the contrary, a recent study \cite{Karolak15} using dynamical mean-field theory (DMFT) discussed the role of a temperature-dependent Hund's exchange energy and showed that the spin-state transition can be driven purely by electronic means through charge and spin fluctuations. In this scheme, a large Coulomb repulsion will suppress charge fluctuations. Taking a cue from these studies, we explored the role of a change in the on-site Coulomb energy $U_{dd}$ and the Co $3d$ - O $2p$ hybridization strength $pd\sigma$
as an electronic alternative playing the role of the breathing lattice type distortion.
Since the e$_g$ electron occupancy increases with the increase of temperature, we can expect an increase in the $pd\sigma$ strength (or occupied bandwidth $W$) and a decrease in the on-site $U_{dd}$ due to screening by itinerant $e_g$ electrons \cite{Solovyev96}.
We found that the observed energy shift ${\Delta}E=40$~meV at 150~K can be explained if  the effective Coulomb energy is reduced by about 0.5~eV and the magnitude of $pd\sigma$ is increased by 0.05~eV \cite{S4}. At 450~K, the energy shift ${\Delta}E$ is 90~meV, consistent with the calculations by using $U_{dd}$=5.5 eV and $pd\sigma= -1.9$ eV, i.e. $U_{dd}$  reduced by 1 eV and $pd\sigma$ shifted by 0.1 eV.  This suggests a further reduction in effective Coulomb energy and also an increase in covalency, leading to the stabilisation of the metallic IS phase due to an avalanche process at higher temperatures. 

In summary, our data and charge-transfer multiplet calculations indicate that the single-ion picture successfully explains the spin-state evolution and suggest an orbital-selective Coulomb energy for the $t_{2g}$ and $e_{g}$ states. Our findings show that the spin-state transition of LaCoO$_3$ is coupled to changes in orbital-selective Coulomb correlations which control spin-charge excitations. In addition, the $e_g$ screening gradually increases with increase of temperature, resulting in a $U/W$ change  associated with the spin-state crossover which leads to a Mott-type insulator-to-metal transition.

%\section*{Acknowledgements}
We thank the NSRRC staff for technical help during our RIXS measurements at the AGM-AGS beamline of Taiwan Light Source. We thank A. Fujimori, G. Y. Guo, M. W. Haverkort, J. H. Park, I. Solovyev, and L. H. Tjeng  for valuable discussions. This work was supported in part by the Ministry of Science and Technology of Taiwan under Grant No. 103-2112-M-213-008-MY3, and by MEXT \& JSPS KAKENHI (JP17H06137 \& JP15H03692) of Japan.

K.T. and J.O. contributed equally to this work.

\end{document}